# Pinning the conformation of a protein (CorA) in a solute matrix with selective binding

Warin Rangubpit[1], Sunan Kitjaruwankul[2], Pornthep Sompornpisut[1], R.B. Pandey[3*]

[1]Center of Excellence in Computational Chemistry, Department of Chemistry, Chulalongkorn University, Bangkok 10330, Thailand
[2]Faculty of Science at Sriracha, Kasetsart University Sriracha Campus, Chonburi 20230, Thailand
[3]Department of Physics and Astronomy, University of Southern Mississippi, Hattiesburg, MS 39406, USA
*Corresponding author, email: ras.pandey@usm.edu

**Abstract:** Conformation of a protein (CorA) is examined in a matrix with mobile solute constituents as a function of solute-residue interaction strength ($f$) by a coarse-grained model with a Monte Carlo simulation. Solute particles are found to reach their targeted residue due to their unique interactions with the residues. Degree of slowing down of the protein depends on the interaction strength $f$. Unlike a predictable dependence of the radius of gyration of the same protein on interaction in an effective medium, it does not show a systematic dependence on interaction due to pinning caused by the solute binding. Spread of the protein chain is quantified by estimating its effective dimension ($D$) from scaling of the structure factor. Even with a lower solute-residue interaction, the protein chain appears to conform to a random-coil conformation ($D \sim 2$) in its native phase where it is globular in absence of such solute environment. The structural spread at small length scale differs from that at large scale in presence of stronger interactions: $D \sim 2.3$ at smaller length scale and $D \sim 1.4$ on larger scale with $f = 3.5$ while $D \sim 1.4$ at smaller length scale and $D \sim 2.5$ at larger length scales with $f = 4.0$.

## 1 Introduction

Understanding the structure of proteins has been a subject of intense interest for decades with an exponential growth in number of publications particularly using Molecular Dynamics simulations. Despite enormous efforts, predicting the nature of 'protein folding' remains a challenging problem [1]. Addressing how proteins reach its equilibrium configurations leads to more questions than answers. For example, what is an equilibrium configuration (minimization of global energy, approaching a steady-state radius of gyration, etc.)? Is equilibrium configuration necessary for a protein to perform its functions? The complexity in predicting the structural properties arises due to competition between the interaction energy and thermal agitation.

The characteristics of an interacting matrix in which the protein is embedded affects the dynamics and structure of the protein. For example, Duncan et al. [2] have recently reported sub-diffusion of proteins and lipids and effects of crowding on their sub-diffusive dynamics in general. Zeindlhofer and Schroeder [3] have reviewed the analysis of biomolecules in aqueous ionic liquid mixtures where they have pointed out that the viscosity of the imbedding matrix affect the structure of the protein. They have cautioned that such effects may be specific to individual protein and therefore should be tested before generalizing the main observation. In this article we focus on a specific protein (CorA) embedded in an interaction-specific matrix with mobile constitutive elements called solutes.

The transmembrane protein CorA consists of 351 residues with a well-defined inner (iCorA) and outer (oCorA) segments in cells [4-10]. Numerous elements in a cell such as lipid molecules, ions, osmolytes, different types of proteins, etc. constitute the crowded environment which delicately control the structure of CorA for selective transport of magnesium ions across the membrane in open and closed states of the ion-channel. Obviously, it is not feasible to incorporate each constitutive elements at once in order to assess its specific effects on different segments of CorA protein. However, it is possible to represent the solutes by simplified particles with appropriate interactions, suitable to capture their specificity. The stochastic movements of each residue (leading to a collective movement of the protein chain) in an effective dynamic matrix environment may provide chances for each residue to interact with other residues of the protein chain. Attractive interactions between some residues (e.g. hydrophilic and electrostatic) and solute may lead to their transient binding while repulsive interaction between other residues may enhance the self-organizing segmental structures (globular or fibrous). Note that the strength of both attractive and repulsive interactions vary from one residue to another. The interplay between the solute-residue interaction and thermal agitation may lead to unique structural evolution. This article is focused on the effect of the strength of interaction on the selective binding of residues with the mobile solute particles and the structural evolution of CorA at a low (native phase) and a high (denatured phase) temperature [11].

It is worth pointing out that we have already examined the structure and dynamics of CorA in absence of environmental complexity and found interesting thermal response of inner and outer segments of the protein [10, 11]. For example, the thermal response of the inner segment shows a continuous transition from globular to random-coil structure on raising the temperature while the outer segment exhibits an abrupt (nearly discontinuous) thermal response in a narrow range of temperature. Unlike in denatured phase, the conformation of the inner segment contracts on raising the temperature in its native phase where the outer segment appears less organized [11]. In an implicit effective medium, the size of the inner segment of the protein decreases in native phase and increases in denatured phase before reaching saturation with the residue-matrix interaction strength; the outer segment shows opposite response to effective medium [12]. In presence of explicit effective solute matrix the protein structure is pinned without a systematic trend with the interaction strength due to rapid binding of solutes with the selective residues as discussed in this article. The model is introduced in the next section followed by results and discussion and a concluding remark.

**2 Model and method**

A bond-fluctuation mechanism [13] is used to model the protein [10-12], a chain of 351 nodes, each representing unique specificity of corresponding residue. The simulation is performed on a cubic lattice with ample degrees of freedom for each residue to perform its stochastic moves and corresponding peptide bonds to fluctuate. Simulation box consists of a protein chain and a large number of solute particles (with a volume fraction c) where the size of a solute is the same as that of a residue [14]. Initially the protein chain is placed in a random configuration and the explicit solute particles are distributed randomly. Each residue and solute molecule interacts with surrounding residues and solute molecules within a range ($r_c$) with a generalized Lennard-Jones potential,

$$U_{ij} = \left[ |\varepsilon_{ij}| \left( \frac{\sigma}{r_{ij}} \right)^{12} + \varepsilon_{ij} \left( \frac{\sigma}{r_{ij}} \right)^{6} \right], \; r_{ij} < r_c \qquad (1)$$

where $r_{ij}$ is the distance between the residues at site $i$ and $j$ or between the residue at site $i$ and residue at site $j$; $r_c = \sqrt{8}$ and $\sigma = 1$ in units of lattice constant. The range of interaction includes ample number of lattice sites that can be occupied by solute molecules or residues of the protein. The degrees of freedom can be further enhanced with finer-grain representations of each residue. A knowledge-based interaction matrix [14-16] is used for the residue-residue pair interaction ($\varepsilon_{ij}$), which is derived from an ensemble of a large number of protein structures from the protein data bank (PDB). The strength $\varepsilon_{ij}$ of the potential (*210* independent elements of a *20 × 20* matrix for *20* amino acids) is unique for each interaction pair with appropriate positive (repulsive) and negative (attractive) values [15, 16].

A solute-solute and solute-residue interactions require 21 unique parameters [14] to capture its specificity in an aquatic solute environment, as considered here for simplicity. The solute at a site (*i*) interacts with a residue at another site (*j*) with a similar interaction as (1) but with $\varepsilon_{ij} = f \varepsilon_r A_{h/p/e}$ which is based on the hydropathy index ($\varepsilon_r$) that controls the attractive and repulsive nature of the residue towards the solute. The hydropathy index is binned into a repulsive interaction ($\varepsilon_r = 0.1$) for all hydrophobic (*H*) residues, attractive interaction ($\varepsilon_r = -0.2$) for all polar (*P*) residues, and more attractive ($\varepsilon_r = -0.3$) for all electrostatic (*E*) residues. The magnitudes of the hydropathy index $\varepsilon_r$ are further weighted with a factor $A_{h/p/e}$ to capture the specificity of each residue within each group (H, P, E). The empirical parameter *f* is introduced to modulates the interaction strength. For example, the interaction of a solute with a hydrophobic residue such as valine (V), $\varepsilon_{ij} = f(0.1)0.933$, the interaction with a hydrophilic residue, say Tryptophan (W) $\varepsilon_{ij} = f(-0.2)0.257$, and the interaction with an electrostatic residue such as lysine (K) $\varepsilon_{ij} = f(-0.3)0.867$. The solute-solute interaction is ignored ($\varepsilon_{ij} = 0$) apart from their excluded volume effect. In the figures and text the interaction strength *f* and *fw* are used interchangeably.

Each residue and solute particle executes their stochastic motion with the Metropolis algorithm. Attempt to move each residue and solute particle once defines unit Monte Carlo step time. Simulations are carried out for sufficiently long time to generate conformational ensembles in steady-state at a low and high temperature regimes each with 5-10 independent samples to analyze a number of local and global physical quantities. Interaction strength *f* is varied. Different lattice sizes are used to make sure that the qualitative trends are independent of the sample size. The results presented here are based on data generated on a $350^3$ sample which provides ample sampling at long-time scales without using excessive computer resources. Reduced units are used for temperature, time step, and spatial length scales in this simulation since our focus is on changes in physical quantities in response to changing the solute interaction strength (*f*) that affect the preferential binding.

**3 Results and discussion**

Solute particles interact with each residues with unique attractive and repulsive interactions controlled by the interaction strength (*f*). Each residue and solute particle perform their stochastic motion. Solute particles are generally more mobile (at least initially) than the residues as their mobility is constrained by the peptide (covalent) bonds. Thus the probability for the solute particle to reach attractive residue and stay there within the range of interaction of the target site for a longer time is higher for higher *f*. Representative snapshots of protein and the solute particles that bind to specific residues are presented in figure 1. We see that the number of solute particles that bind to residue increases with the solute-residue interaction strength (*f*). Further, there is no appreciable change in size of the protein except some variations in segmental organization. It is therefore difficult to quantify overall changes in size of the protein in response to solute-residue interactions.

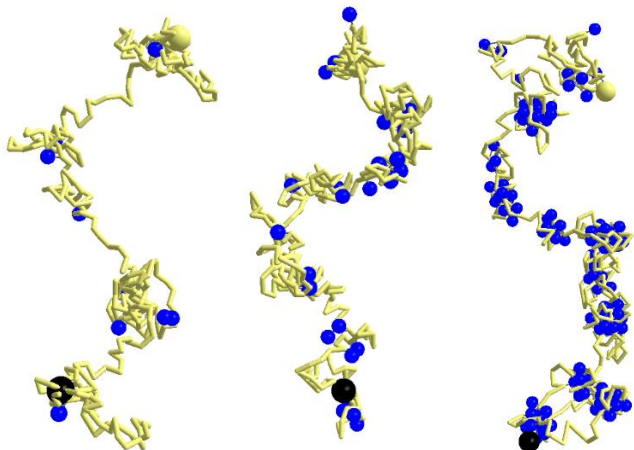

Figure 1: Snapshots of the protein chain CorA at the end of a million time step. Bonds along the backbone of the protein contour are in yellow where the first residue is represented by the large black and the last residue by yellow sphere; solute particles within the range of interaction of each residue are shown in blue dots. The interaction strength $f = 2.0, 2.5,$ and $3.5$ from left to right at T = 0.020.

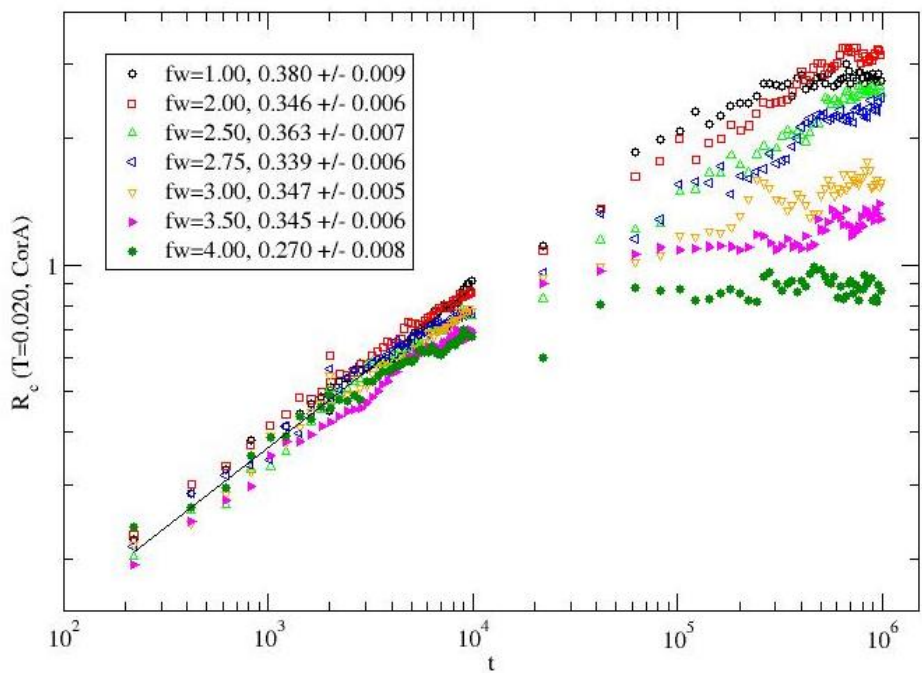

Figure 2: Root mean square (RMS) displacement ($R_c$) of the center of mass of the protein chain with the time step ($t$) for a range of solute-residue interaction $f = 1.00 – 4.00$ at a low temperature $T = 0.020$. The power-law exponent $\nu$ ($R_c \propto t^\nu$) in the short time regime ($t \sim 10^2 – 10^4$) is also included for each interaction strength.

How does the protein chain move as the residues attempt to perform their stochastic motion? One may be able to assess the protein dynamics by examining the variation of the root mean square (RMS) displacement ($R_c$) of the center of mass of the protein chain with the time step as shown in figure 2. As the protein chain moves, its RMS displacement increases with the time step. Note that the movement is faster in short time (initially), followed by a slowdown asymptotically in long time regime. Asymptotic slowdown depends on the solute-residue interaction strength (*f*); the higher the interaction, faster is the approach from very slow movement (*f = 1.00 – 2.75*) to almost standstill (*f = 3.00 – 4.00*). The dynamics of the protein chain in short time regime can be assessed by a power law dependence of its RMS displacement, i.e., $R_c \propto t^\nu$, where the exponent $\nu$ characterizes the nature of dynamics. We see that the dynamics of the protein is sub-diffusive $\nu < ½$ for entire range of solute-residue interaction in the short time regime. In asymptotic regime the dynamics of the protein is not only sub-diffusive with low interactions but almost vanishes ($\nu \sim 0$) with stronger interactions (*f = 3.00 – 4.00*). Slowing down of the protein chain occurs as the solute particles bind to their target residues; the overall motion ceases as the conformation of the protein is pinned down by ample binding as seen in figure 1.

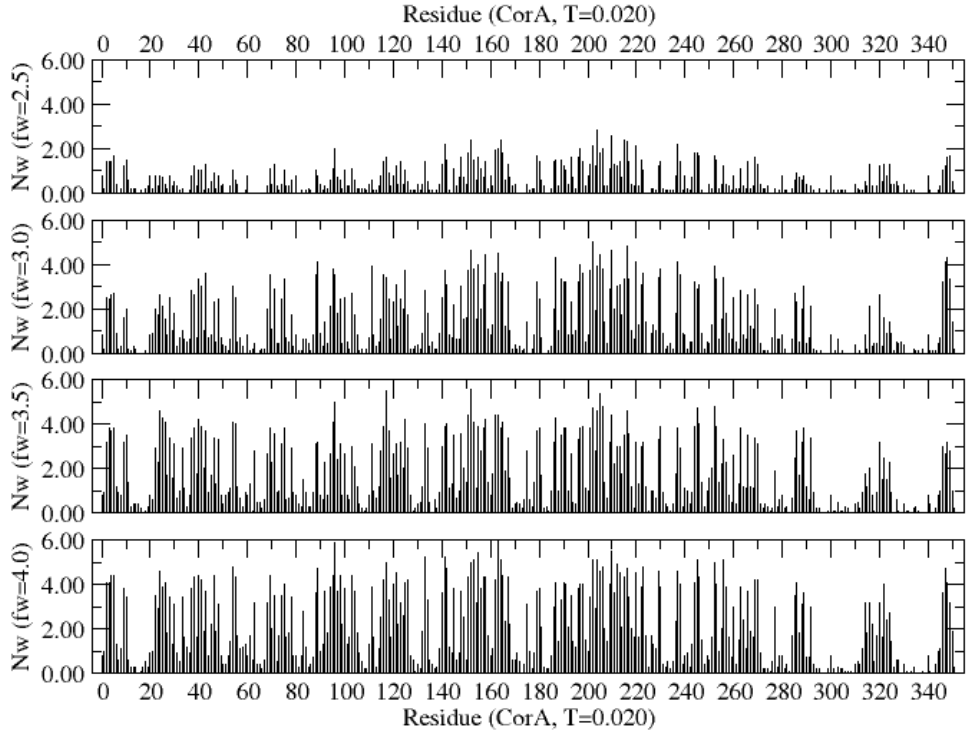

Figure 3: Binding profile (average number $N_w$ of solute around each residues) of CorA for solute-residue interaction *f=2.5, 3.0, 3.5, 4.0* at a temperature *T = 0.020*.

In order to assess the extent of binding and identify the targeted residues, let us examine the binding profile, i.e. the average number ($N_w$) of solute particles around each residue within the range of interaction. Figure 3 shows the average number of solute particles around each residue for a range of solute-residue interaction strength *f=2.5, 3.0, 3.5, 4.0* at a temperature *T = 0.020*. Outer and inner transmembrane segments of the protein CorA consists of $^1$M $^2$E ….$^{291}$M and $^{292}$K $^{293}$V … $^{351}$L. The extent of binding is enhanced with stronger solute-residue interaction. A large fraction of outer segment of CorA residues are pinned by high degree of solute binding specially at stronger interactions (*f = 4.00*) while a considerable fraction (about half) of inner transmembrane segments remain unbounded (i.e. $^{292}$K $^{293}$V … $^{309}$G, $^{330}$V $^{331}$L … $^{343}$V). The inner segment of the protein CorA is more mobile than that of the outer transmembrane segment and perhaps more responsive in self-assembly as recently observed [17].

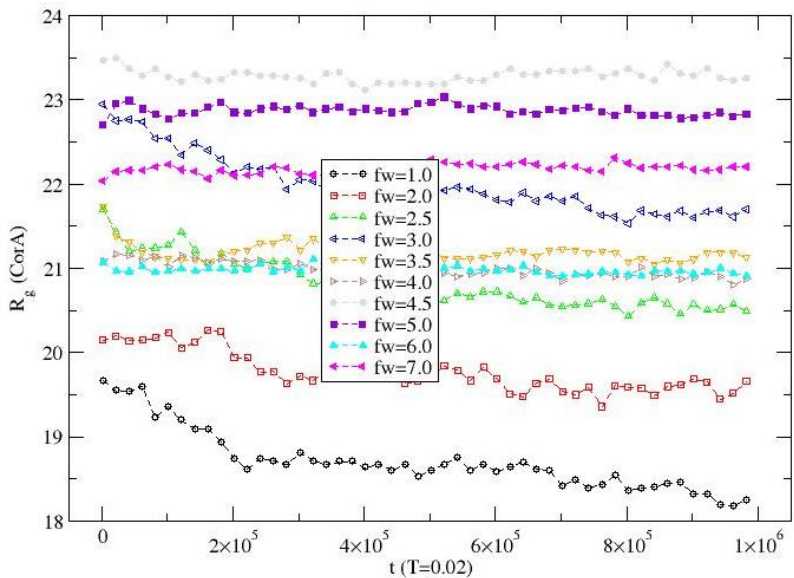

Figure 4: Radius of gyration of CorA versus time step for solute-residue interaction *f=1.0 − 7.0* at *T = 0.02*.

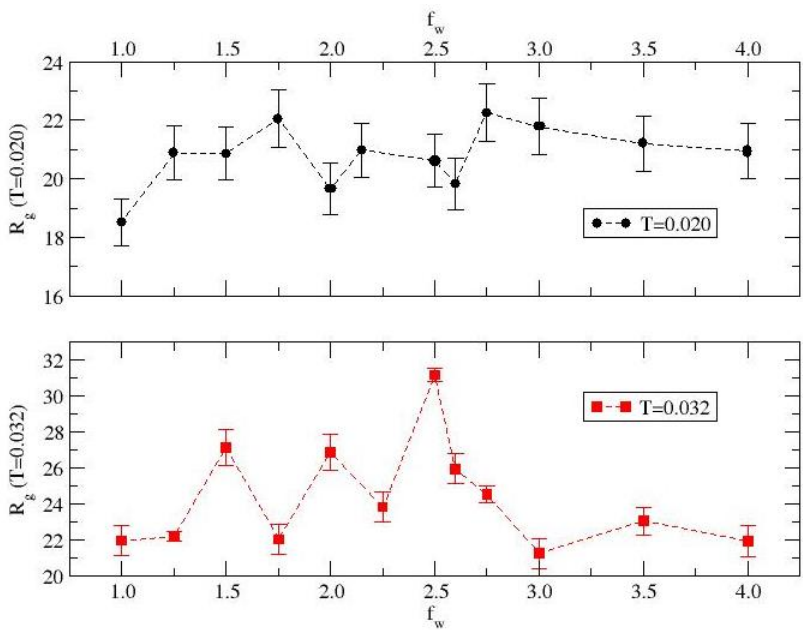

Figure 5: Radius of gyration of the protein (CorA) versus solute-residue interaction strength *f ≡ fw* at temperature *T = 0.020, 0.032*.

Because of the selective binding of solute particles, the protein conformation is arrested in a relatively short time step. Figure 4 shows the variation of the radius of gyration with the time step for a wide range of solute-residue interaction (*f=1.0 − 7.0*) at a low temperature. We see that the radius of gyration becomes reaches its steady-state rather fast particularly at higher values of *f* and that the binding enhances the stabilization. The variation of the average radius of gyration with the interaction strength (figure 5) shows almost no systematic dependence on the interaction at temperatures T =0.020, 0.032 despite large fluctuations. Lack of a trend is due to pinning of the conformations of the protein as the solute particles move fast and bind the targeted residue. Note that lack of a systematic dependence on the interaction with the underlying matrix due to pinning is different from that of the protein in an effective medium [13] where the inner transmembrane segment exhibits a systematic dependence in both native and denatured phases.

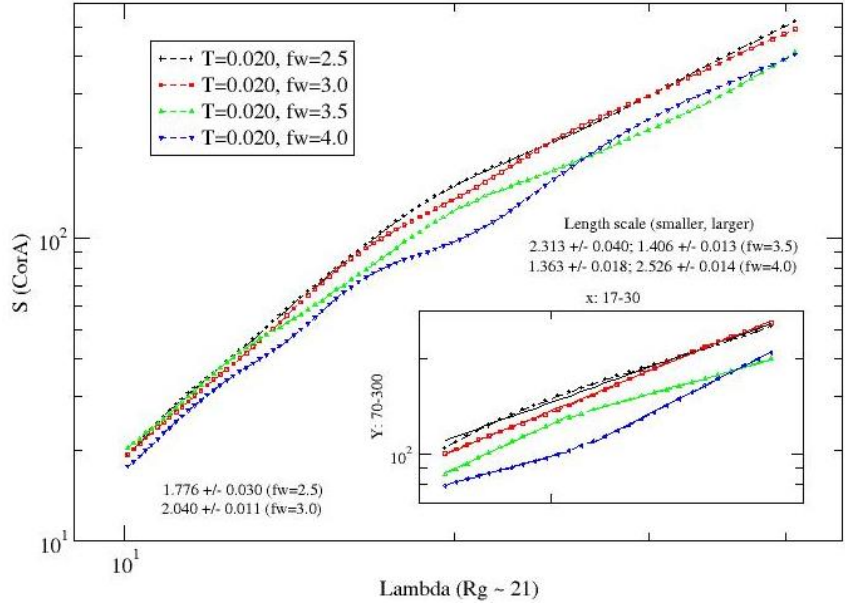

Figure 6: Structure factor (S) of the protein versus wave length (Lambda) for the interaction strength *fw = 2.5, 3.0, 3.5,* and *4.0* at *T = 0.020*. The inset is amplified version within a limited wavelength range spanning over the magnitude of the radius of gyration ($R_g \sim 21$). The slope ($1/\gamma = D$) is a measure of the effective dimension (*D*) of the protein which is about *2* with lower values of *fw = 2.5, 3.0* but varies with length scales at higher values of *fw = 3.5, 4.0*.

To quantify the conformational spread of the protein, we analyze the structure factor S(q),

$$S(q) = \langle \frac{1}{N} \left| \sum_{j=1}^{N} e^{-i\vec{q}\cdot r_j} \right|^2 \rangle_{|\vec{q}|} \qquad (2)$$

where $r_j$ is the position of each residue in all protein chains and $|q| = 2\pi/\lambda$ is the wave vector of wavelength $\lambda$. Using a power-law scaling of the structure factor with the wave vector, i.e.,

$$S(q) \propto q^{-1/\gamma} \qquad (3)$$

one may study the spread of residues over the length scale $\lambda$ by evaluating the exponent $\gamma$ which describes the mass (residue) distribution. The slope $D = 1/\gamma$ of the data set spanning over the length scale comparable to radius of gyration is a measure of the effective dimension (*D*) of the protein. At lower values of solute-residue interaction *f = 2.5, 3.0*, the conformation of the protein chain is like a random-coil ($D \sim 2$) at $T=0.020$ where it conform to a globular structure in absence of solute. Effective dimension of the protein depends on the length with stronger solute-residue interaction. For example, $D \sim 2.3$ at smaller length scale and $D \sim 1.4$ (very linear structure) on larger scale with $f = 3.5$. Reverse is the case with stronger interaction $f = 4.0$, where $D \sim 1.4$ at smaller length scale and $D \sim 2.5$ at larger length scales. Structural variability suggests that a wide range of fluctuating conformations can be pinned by appropriate solute.

## 4 Conclusions

Conformational dynamics of a protein (CorA) in an interacting matrix with mobile solute particles is examined as a function of solute-residue interaction by a coarse-grained Monte Carlo simulation. Initially the protein chain is placed in the simulation box in a random configuration in presence of a random distribution of solute particles. Each residue and solute particle perform their stochastic movement. Because of the higher mobility of solute particles and specificity of the solute-residue interactions, they reach their specific target rather fast. As a result the conformation of the protein is constrained by the solute particles that bind to specific protein sites and pinned with stronger solute-residue interactions.

The protein chain exhibits different dynamics in short and long time regime: the short time dynamics is sub-diffusive and not very sensitive to solute-residue interaction strength (fw), the long time (asymptotic) dynamics depends strongly on the interaction strength. The protein chain becomes almost immobile as solute particles bind to their target residues at higher solute-residue interactions. Lack of a systematic trend in conformational response to solute-residue interaction seems to occur due to pinning down a vast number of specific residues, particularly in the outer transmembrane segment of the protein. This observation is very different from that of the same protein chain in an effective medium [12] where there is a systematic dependence of the protein conformation on the residue-matrix interaction.

Spread of the protein chain can be quantified by estimating its effective dimension from scaling of the structure factor. In a relatively lower solute-residue interaction, the protein chain appears to conform to a random-coil conformation at a low temperature where it is globular in absence of such solute environment; this is obviously due to some degree of pinning of the protein conformation. Presence of stronger interacting solute leads to higher degree of pinning down the conformations that shows different spread at lower and larger length scales. For example, $D \sim 2.3$ at smaller length scale and $D \sim 1.4$ (very linear fibrous structure) on larger scale with the solute-residue interaction strength $f = 3.5$. Reverse is the case on increasing the interaction strength $f = 4.0$, where $D \sim 1.4$ at smaller length scale and $D \sim 2.5$ at larger length scales. Thus by selecting interacting matrix with specific solute-residue interaction, one may be able to achieve a desirable conformation of the protein.

Acknowledgement: This research has been supported by the Ratchadaphiseksomphot Endowment Fund, Chulalongkorn University to PS, the Chulalongkorn university dusadi phipat scholarship to WR. Support from Chulalongkorn University for the visiting professorship is gratefully acknowledged by RBP. The authors acknowledge HPC at The University of Southern Mississippi supported by the National Science Foundation under the Major Research Instrumentation (MRI) program via Grant # ACI 1626217.